\documentclass[english,reprint,superscriptaddress]{revtex4-2}
\usepackage[T1]{fontenc}
\usepackage[latin9]{inputenc}
\usepackage{babel}
\usepackage{array}
\usepackage{bm}
\usepackage{amsmath}
\usepackage{amssymb}
\usepackage{graphicx}
\usepackage{hyperref}
\hypersetup{
 colorlinks=true,allcolors=blue,
 breaklinks=true}

\makeatletter

\providecommand{\tabularnewline}{\\}

\makeatother

\begin{document}
\title{Active oscillatory associative memory}
\author{Matthew Du}
\affiliation{Department of Chemistry, University of Chicago, Chicago, Illinois 60637, USA}
\affiliation{The James Franck Institute, University of Chicago, Chicago, Illinois 60637, USA}
\author{Agnish Kumar Behera}
\affiliation{Department of Chemistry, University of Chicago, Chicago, Illinois 60637, USA}
\author{Suriyanarayanan Vaikuntanathan}
\affiliation{Department of Chemistry, University of Chicago, Chicago, Illinois 60637, USA}
\affiliation{The James Franck Institute, University of Chicago, Chicago, Illinois 60637, USA}
\date{\today}
\begin{abstract}
Traditionally, physical models of associative memory assume conditions of equilibrium. 
Here, we consider a prototypical oscillator model of associative memory and study how active noise sources that drive the system out of equilibrium, as well as nonlinearities in the interactions between the oscillators, affect the associative memory properties of the system. Our simulations
show that pattern retrieval under active noise is more robust to the
number of learned patterns and noise intensity than under passive
noise. To understand this phenomenon, we analytically
derive an effective energy correction due to the temporal correlations
of active noise in the limit of short correlation decay time. We find
that active noise deepens the energy wells corresponding to the patterns
by strengthening the oscillator couplings, where the more nonlinear
interactions are preferentially enhanced. Using replica theory, we
demonstrate qualitative agreement between this effective picture and
the retrieval simulations. Our work suggests that nonlinearity in
the oscillator couplings can improve memory under nonequilibrium conditions.
\end{abstract}
\maketitle
An essential feature of biological organisms is memory, which involves
the learning and retrieval of information. The importance of this
cognitive ability has prompted many studies of memory from a physics
perspective \citep{keim_memory_2019,sokolov_towards_2021}, specifically
associative, or content-addressable, memory \citep{amit_modeling_1989,nakano_associatron-model_1972,kohonen_correlation_1972,amari_learning_1972}. 

A well explored model of associative memory is the Hopfield neural
network \citep{hopfield_neural_1982}, which resembles a fully connected
Ising model with nonuniform and random interaction strengths. The
neurons are modeled as spins and the learned patterns, which are
drawn from a probability distribution, are encoded in the couplings
between spins. Naturally, the memory capabilities of the Hopfield
model can be understood from the perspective of statistical mechanics
\citep{amit_statistical_1987,amit_storing_1985,feigelman_statistical_1986,coolen_bookchapter1_2001}.
Refs. \citep{amit_statistical_1987,amit_storing_1985,feigelman_statistical_1986}
showed that thermal fluctuations in the firing rates of neurons leads
to poorer recall of patterns. 

However, it is unclear whether such findings, which characterize the
system at equilibrium, apply to biological systems which are expected to be subjected to noise sources that do not satisfy the constraints of equilibrium. Indeed, a proper physical description of living
systems requires nonequilibrium aspects \citep{gnesotto_broken_2018,fang_nonequilibrium_2019}.
Notably, the flocking of birds \citep{vicsek_novel_1995,toner_hydrodynamics_2005}
and the dynamics of the cytoskeleton \citep{julicher_active_2007,banerjee_actin_2020}
are modeled as active matter \citep{ramaswamy_mechanics_2010,marchetti_hydrodynamics_2013,bechinger_active_2016},
particles that consume energy from the environment to power their
movement. Some active materials, e.g., epithelial cells \citep{sepulveda_collective_2013}
and colloids in a bacterial solution \citep{maggi_generalized_2014},
can be described as particles subjected to so-called active noise
\citep{fodor_how_2016,martin_statistical_2021}, which has temporal
correlations (i.e., is colored noise) and does not obey detailed balance
with respect to the damping. In fact, the membrane potential of 
neurons can have temporally correlated fluctuations \citep{chacron_suprathreshold_2000,fourcaud_dynamics_2002}.

Additionally, a feature that the Hopfield model does not exhibit is the oscillatory
activity of neurons \citep{selverston_oscillatory_1985,ward_synchronous_2003,wang_neurophysiological_2010}.
This motivation led to variants \citep{schuster_model_1990-1,arenas_phase_1994,fukai_memory_1994,fukai_model_1994,park_synchronization_1995,aonishi_phase_1998,aoyagi_retrieval_1998}
of the Kuramoto model \citep{kuramoto_self-entrainment_1975} of nonlinearly
coupled oscillators, whose Hopfield-like interactions give rise to
associative memory. These analog systems, due to their simplicity,
can be studied in much detail using analytical methods \citep{arenas_phase_1994,fukai_memory_1994,aoyagi_retrieval_1998}.

Here, we study an oscillatory Hopfield network in the presence of
active noise. Inspired by the Kuramoto-type models of \citep{fukai_memory_1994,nishikawa_capacity_2004},
the system here has not been studied before, to the best of our knowledge,
even in the absence of noise. 
Our numerical simulations reveal that the recall 
of information is enhanced under active noise,
relative to under comparable equilibrium conditions. 
We then analytically derive, in a perturbative limit, an effective energy correction associated with
the temporal correlations of active noise \citep{fodor_how_2016,maitra_enhanced_2020,jung_dynamical_1987,maggi_multidimensional_2015}. 
Using replica theory \citep{mezard_spin_1987}, we find that the effective
picture describes qualitatively the improvement of associative
memory under active noise. Our work sheds light on how nonlinearity
in the neuron couplings impacts memory under nonequilibrium conditions. 

The paper is structured as follows. 
We introduce the Kuramoto-type Hopfield model 
in Sec. \ref{sec:Model}.
Then, we numerically simulate pattern retrieval 
in Sec. \ref{sec:simulations}.
Sec. \ref{sec:Theory} describes the theoretical analysis,
in which we derive the effective Hamiltonian 
(Sec. \ref{subsec:effective-equilibrium}) 
and apply replica theory to it (Sec. \ref{subsec:Replica-theory}). 
We conclude in Sec. \ref{sec:conclusion}. 

\section{Model\label{sec:Model}}

\begin{figure*}
\centering\includegraphics[width=6.6in]{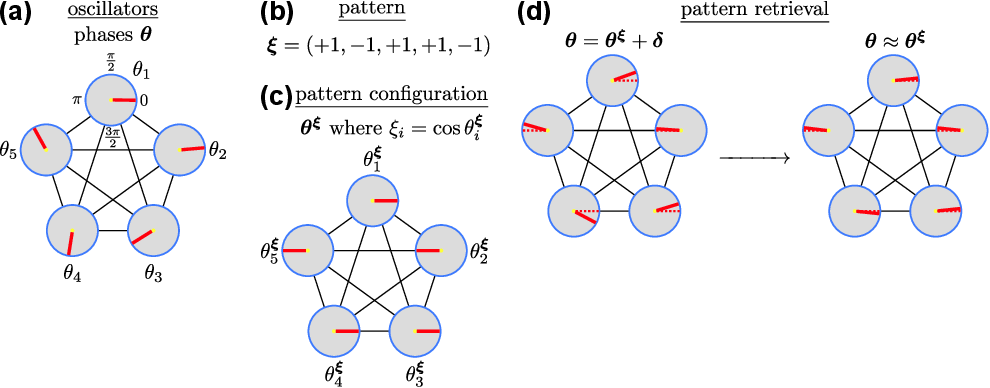}

\caption{Oscillatory associative memory. (a) $N$ coupled oscillators, represented
as a graph whose vertices are the oscillator phases $\bm{\theta}$
and the edges are the couplings. Here, $N=5$. (b) A pattern, which
is a vector $\bm{\xi}\in\{\pm1\}^{N}$. Here, $\bm{\xi}=(+1,-1,+1,+1,-1)$.
(c) A pattern configuration, which for pattern $\bm{\xi}$ is any
$\bm{\theta}^{\bm{\xi}}$ such that $\cos\theta_{i}^{\bm{\xi}}=\xi_{i}$.
Here, $\bm{\theta}^{\bm{\xi}}=(0,\pi,0,0,\pi)$. (d) Pattern retrieval.
The system starts slightly displaced from a target pattern $\bm{\xi}$
($\bm{\theta}=\bm{\theta}^{\bm{\xi}}+\bm{\delta}$ for nonzero $\bm{\delta}$)
and then approaches the pattern ($\bm{\theta}\approx\bm{\theta}^{\bm{\xi}}$).
\label{fig:oam}}

\end{figure*}

We consider $N$ periodic oscillators whose interactions endow the
system with associative memory (Fig. \ref{fig:oam}a). Specifically,
we focus on the Hamiltonian
\begin{align}
H & =-\frac{1}{2N}\sum_{\mu=1}^{p}\sum_{i,j=1}^{N}\xi_{i}^{\mu}\xi_{j}^{\mu}\cos\theta_{i}\cos\theta_{j}\nonumber \\
 & \quad-\frac{\epsilon}{4N}\sum_{i,j=1}^{N}\cos2\theta_{i}\cos2\theta_{j},\label{eq:h}
\end{align}
where $\theta_{1},\dots,\theta_{N}$ are the oscillator phases. Analogous
in form to the Hamiltonian of \citep{nishikawa_capacity_2004}, $H$
is a generalization of the Kuramoto model \citep{kuramoto_self-entrainment_1975}
that features associative memory. Collectively, the phases and their
cosines will be referred to as $\bm{\theta}=(\theta_{1},\dots,\theta_{N}$)
and $\mathbf{c}(\bm{\theta})=(\cos\theta_{1},\dots,\cos\theta_{N})$,
respectively. 

The first term of $H$ encodes $p$ patterns (Fig. \ref{fig:oam}b)
$\bm{\xi}^{1},\dots,\bm{\xi}^{p}\in\{\pm1\}^{N}$ learned by the Hebb
rule \citep{hebb_organization_2005,hopfield_neural_1982}. Appearing
in the model of Ref. \citep{fukai_memory_1994}, this term is analogous
to the Hopfield Hamiltonian \citep{hopfield_neural_1982}, where the
variable of spin $i$ is replaced by $\cos\theta_{i}$ and the couplings
are of the form $\cos\theta_{i}\cos\theta_{j}$. All pattern elements
are independent and identically distributed (i.i.d.) random variables
drawn uniformly from $\{\pm1\}$ (i.e., a binomial distribution).
When the pattern loading $\alpha=p/N$ is sufficiently small, the
system has energy minima near the pattern configurations (Fig. \ref{fig:oam}c)
$\{\bm{\theta}^{\bm{\xi}^{\mu}}\}_{\mu=1}^{p}$ , where $\mathbf{c}(\bm{\theta}^{\bm{\xi}^{\mu}})=\bm{\xi}^{\mu}$.
Henceforth, we will sometimes write ``pattern'' as shorthand for
``pattern configuration.'' 

The second term of $H$ enhances the stability of the learned patterns,
as well as all other patterns, whose configurations also satisfy $\mathbf{c}(\bm{\theta})\in\{\pm1\}^{N}$.
Characterized by strength $\epsilon$, this term has couplings of
the form $\cos2\theta_{i}\cos2\theta_{j}=(2\cos^{2}\theta_{i}-1)(2\cos^{2}\theta_{j}-1)$,
making it more nonlinear than the pattern-encoding term. By varying
$\epsilon$ in our calculations below, we can understand how nonlinearity
affects memory in the presence of noise. We will restrict ourselves
to intermediate values of $\epsilon$, as the learned patterns are
further stabilized (compared to $\epsilon=0$) while the patterns
that have not been learned remain relatively unstable \citep{nishikawa_capacity_2004}. 

In pattern retrieval, the system starts near a pattern $\bm{\xi}\in\{\pm1\}^{N}$
and then moves towards it (Fig. \ref{fig:oam}d).
The initial position is $\bm{\theta}(0)=\bm{\theta}^{\bm{\xi}}+\bm{\delta}$,
where $\mathbf{c}(\bm{\theta}^{\bm{\xi}})=\bm{\xi}$, $\bm{\delta}$
is a vector of $N$ i.i.d. random variables drawn from $\mathcal{N}(0,\sigma^{2})$,
and $\sigma$ determines the typical size of the initial offset. 
To represent neurons isolated from external forces,
one would evolve the system via gradient descent
on the energy landscape of Hamiltonian (\ref{eq:h}). 
However, real neurons reside in a biological environment,
which can induce nonequilibrium fluctuations 
in the firing rates. 
Thus, we evolve the system 
according to the dynamics of active Ornstein-Uhlenbeck
particles \citep{fodor_how_2016,martin_statistical_2021}: 
\begin{equation}
\frac{d\bm{\theta}}{dt}=-\nabla_{\bm{\theta}}H+\bm{\eta}(t).\label{eq:eom}
\end{equation}
The noise $\bm{\eta}$ is a stationary Gaussian process that satisfies
\begin{align}
\langle\eta_{i}(t)\rangle & =0,\\
\langle\eta_{i}(t)\eta_{j}(t')\rangle & =\frac{T}{\tau}\delta_{ij}e^{-|t-t'|/\tau},
\end{align}
where the temporal correlations exponentially decay with persistence
time $\tau$ and initial value $T/\tau$. For $\tau>0$, the noise
is active,
and the system is kept out of equilibrium. In the limit $\tau\rightarrow0$, the temporal correlations
satisfy
\begin{equation}
\lim_{\tau\rightarrow0}\langle\eta_{i}(t)\eta_{j}(t')\rangle=2T\delta_{ij}\delta(t-t'),
\end{equation}
corresponding to thermal noise which drives 
the system towards equilibrium with temperature $T$. 
Hereafter,
we refer to this type of noise as passive. 
To characterize the proximity
of the system to $\bm{\xi}$, the order parameter that we use is the
overlap $\frac{1}{N}\left[\bm{\xi}\cdot\mathbf{c}(\bm{\theta})\right]$,
which ranges from 0 (farthest from pattern) to 1 (at pattern). After
the system has evolved for a long time, the overlap indicates the
accuracy with which the pattern has been retrieved. 

\section{simulations\label{sec:simulations} }

We simulate pattern retrieval for a system with $N=200$ oscillators.
Without loss of generality, we consider the retrieval of learned pattern
$\mu=1$ ($\bm{\xi}^{1}$), hereafter referred to as ``target pattern.''
We choose $\sigma=0.1$, for which the typical initial overlap with
the target pattern is $\approx0.995$. The system is evolved from
$t=0$ to $t=800$ by numerically integrating equation of motion (\ref{eq:eom})
using a time step of $\Delta t=0.004$. Below, the simulation results
represent an average over 16 pattern sets and 8 initial configurations.
These numbers are high enough such that the error due to finite sampling
can be neglected. 

\begin{figure}
\centering\includegraphics[width=3.3in]{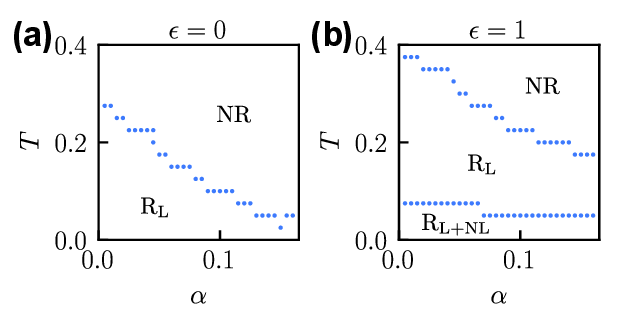}

\caption{Retrieval phase diagram under passive noise, as calculated by simulation,
for stability-enhancing coupling (a) $\epsilon=0$ and (b) $\epsilon=1$.
In the retrieval phase R$_{\text{L+NL}}$, both learned and not-learned
patterns can be retrieved (i.e., with final overlap > 0.8). In the
retrieval phase R$_{\text{L}}$, only learned patterns can be retrieved.
In the no retrieval (NR) phase, no patterns can be retrieved. \label{fig:passive}}
\end{figure}

We first focus on the system with just the pattern-encoding term,
i.e., with $\epsilon=0$ (Fig. \ref{fig:passive}a). In the absence
of noise ($T=0$), the learned patterns can be retrieved at higher
pattern loading ($\alpha$) compared to the Hopfield model, for which
retrieval fails at $\alpha\gtrsim0.138$ \citep{amit_statistical_1987,amit_storing_1985}.
However, the critical pattern loading quickly decays to zero as passive
noise ($T\neq0$, $\tau\rightarrow0$) is added. In fact, the system
is less robust to noise than the Hopfield model, for which patterns
can be retrieved up to $T=1$ \citep{amit_statistical_1987,amit_storing_1985}. 

When the stability-enhancing term {[}second term of Eq. (\ref{eq:h}){]}
is turned on ($\epsilon>0$, Fig. \ref{fig:passive}b), the retrievability
of the learned patterns deteriorates less rapidly with temperature
and pattern loading. In particular, retrieval is enabled at higher
temperatures for which it does not occur when $\epsilon=0$. These
improvements come at the expense of being able to, at lower temperatures,
also retrieve patterns that have not been learned (Fig. \ref{fig:passive}b,
$\text{R}_{\text{L+NL}}$ phase). For this reason, we restrict ourselves
to $\epsilon\leq1$ throughout this work.

\begin{figure}
\centering\includegraphics[width=3.3in]{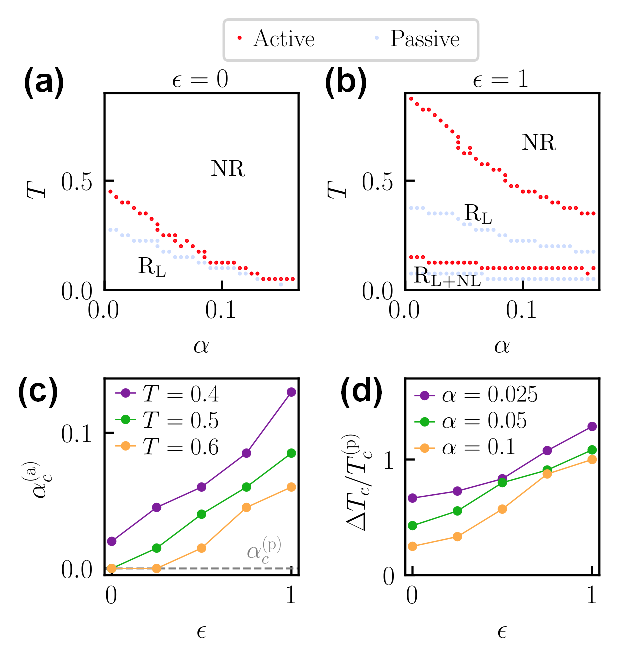}

we \caption{(a, b) Retrieval phase diagram under active noise ($\tau=1$, red),
as calculated by simulation, for stability-enhancing coupling (a)
$\epsilon=0$ and (b) $\epsilon=1$. See the caption of Fig. \ref{fig:passive}
for a description of the various phases. For comparison, the transition
lines under passive noise (reproduced from Fig. \ref{fig:passive})
are shown as blue translucent points. 
(c) Critical pattern loading (for the
$\text{R}_{\text{L}}\rightarrow\text{NR}$ transition) under active
noise ($\alpha_{c}^{(\text{a})}$) as a function of $\epsilon$. Results
are shown for various noise strengths ($T$) at which patterns cannot
be retrieved under passive noise (i.e., $\alpha_{c}^{(\text{p})}=0$,
as indicated by gray dashed line). 
(d) Relative difference ($\Delta T_{c}/T_{c}^{(\text{p})}$,
where $\Delta T_{c}=T_{c}^{(\text{a})}-T_{c}^{(\text{p})}$) in the
critical noise strength (for the $\text{R}_{\text{L}}\rightarrow\text{NR}$
transition) between active ($T_{c}^{(\text{a})}$) and passive ($T_{c}^{(\text{p})}$)
noise as a function of $\epsilon$. Results are shown for various
pattern loadings ($\alpha$).  \label{fig:active}}
\end{figure}

Having characterized the system under passive noise, we now consider
the system under active noise of equal strength ($T$) and with persistence
time $\tau=1$. Compared to passive noise, pattern retrieval is more
resistant to pattern loading and noise strength (Fig. \ref{fig:active},
a-b). Notably, retrieval is successful at noise strengths for which
it fails under passive noise. The critical pattern loading at these
noise strengths grows with $\epsilon$ (Fig. \ref{fig:active}c).
Similarly, for a given pattern loading, active noise raises the critical
noise strength (i.e., the noise strength beyond which retrieval does
not occur), and the relative increase scales with $\epsilon$ (Fig.
\ref{fig:active}d). These findings suggest that higher nonlinearity
in the oscillator couplings allows active noise to better stabilize
pattern configurations which are unstable under passive noise.

\section{Theory\label{sec:Theory}}

To gain further insight on the improvement of memory due to active
noise, we employ a perturbative expansion with respect to persistence
time $\tau$ \citep{fodor_how_2016,maitra_enhanced_2020}. This approach
yields an effective equilibrium steady state, characterized by temperature
$T$ and a modified Hamiltonian. As we will show below, for the system
studied here, the temporal correlations can be recast as an effective
renormalization of coupling strengths.We then study the associated
effective Hamiltonian using replica theory \citep{mezard_spin_1987}.

\subsection{Effective equilibrium \label{subsec:effective-equilibrium}}

Consider the steady-state probability distribution $P(\bm{\theta})$
of system configurations $\bm{\theta}$ in the limit of small persistence
time, $\tau\ll1$. Following the method of either \citep{fodor_how_2016}
or \citep{maitra_enhanced_2020}, we can write $P(\bm{\theta})$ in
powers of $\tau$ and keep terms up to $O(\tau)$, yielding a Boltzmann-like
distribution $P(\bm{\theta})\propto\exp(-H_{\text{eff}}/T)$, which
is determined by an effective Hamiltonian with the general form \citep{fodor_how_2016}
\begin{equation}
H_{\text{eff}}=H+\tau\left(\frac{1}{2}\left|\nabla_{\bm{\theta}}H\right|^{2}-T\nabla_{\bm{\theta}}^{2}H\right).\label{eq:general-heff}
\end{equation}
Corresponding to an effective equilibrium steady state, this result
can be alternatively obtained using the unified colored noise approximation
\citep{jung_dynamical_1987,maggi_multidimensional_2015}. Plugging
$H$ {[}Eq. (\ref{eq:h}){]} in the Laplacian term gives
\begin{align}
H_{\text{eff}} & =-\frac{\zeta_{1}}{2N}\sum_{\mu=1}^{p}\sum_{i,j=1}^{N}\xi_{i}^{\mu}\xi_{j}^{\mu}\cos\theta_{i}\cos\theta_{j}\nonumber \\
 & \quad-\frac{\zeta_{2}\epsilon}{4N}\sum_{i,j=1}^{N}\cos2\theta_{i}\cos2\theta_{j}\nonumber \\
 & \quad+\frac{\tau}{2}\left|\nabla_{\bm{\theta}}H\right|^{2}\nonumber \\
 & \quad+\tau T\left\{ \frac{1}{N}\sum_{\mu=1}^{p}\left[\bm{\xi}^{\mu}\cdot\mathbf{s}(\bm{\theta})\right]^{2}+\frac{2\epsilon}{N}\left[\mathbf{1}\cdot\mathbf{s}(2\bm{\theta})\right]^{2}\right\} ,\label{eq:heff0}
\end{align}
where $\zeta_{1}=1+2\tau T$, $\zeta_{2}=1+8\tau T$, and $\mathbf{s}(\bm{\theta})=(\sin\theta_{1},\dots,\sin\theta_{N})$.
While it is possible to write $\frac{\tau}{2}\left|\nabla_{\bm{\theta}}H\right|^{2}$
more explicitly in terms of $\bm{\theta}$, we have simply chosen
not to do so, since the current expression is sufficient for physical
interpretation (see below). Previously, we applied this small-$\tau$
expansion to the Hopfield model under active noise \citep{behera_enhancing_2022}.
However, due to the constraint on the norm of the spin vector, we
needed to make additional approximations to arrive at the effective
Hamiltonian. 

Note that the $\frac{\tau}{2}\left|\nabla_{\bm{\theta}}H\right|^{2}$
term of $H_{\text{eff}}$ {[}third line of Eq. (\ref{eq:heff0}){]}
is the squared norm of the (conservative) force ($-\nabla_{\bm{\theta}}H$),
which is zero at the energy minima near the patterns. Similarly, the
fourth term of $H_{\text{eff}}$ {[}fourth line of Eq. (\ref{eq:heff0}){]}
is zero at the patterns. Thus, these last two terms should play a
minor role in pattern retrieval, which is determined by the energy
landscape in the neighborhood of the patterns. 

\begin{figure}
\centering\includegraphics[width=3.3in]{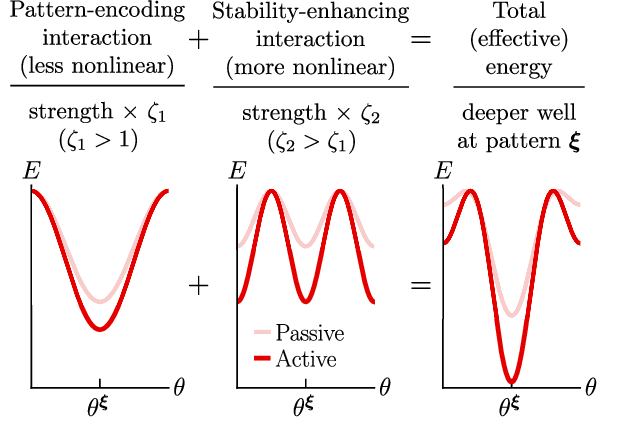}

\caption{Schematic illustration of the mechanism, as described by effective
Hamiltonian $H_{\text{eff}}$ {[}Eq. (\ref{eq:heff}){]}, by which
active noise improves memory compared to passive noise. The temporal
correlations of active noise effectively strengthen the oscillator
interactions, where the more nonlinear interactions are preferentially
enhanced. This strengthening leads to a deeper energy well at the
patterns, and thus their retrieval becomes more robust to noise intensity
and number of learned patterns. \label{fig:h-eff}}

\end{figure}

We therefore expect pattern retrieval under active noise to be approximately
described by the effective Hamiltonian
\begin{align}
H_{\text{eff}} & =-\frac{\zeta_{1}}{2N}\sum_{\mu=1}^{p}\sum_{i,j=1}^{N}\xi_{i}^{\mu}\xi_{j}^{\mu}\cos\theta_{i}\cos\theta_{j}\nonumber \\
 & \quad-\frac{\zeta_{2}\epsilon}{4N}\sum_{i,j=1}^{N}\cos2\theta_{i}\cos2\theta_{j},\label{eq:heff}
\end{align}
which has the same form as the original Hamiltonian {[}Eq. (\ref{eq:h}){]},
except with the pattern-encoding term (first term) scaled by $\zeta_{1}$
and the stability-enhancing term scaled by $\zeta_{2}$. Since $\zeta_{1},\zeta_{2}>1$,
active noise stabilizes the learned patterns (compared to passive
noise) by deepening their respective energy wells. In particular,
as implied by $\zeta_{2}/\zeta_{1}>1$, the stability-enhancing term
is amplified relative to the pattern-encoding term. More generally,
the pattern-stabilizing interactions with more nonlinearity (i.e.,
the stability-enhancing term) are strengthened over those with less
(i.e., the pattern-stabilizing term). This broader view suggests the
importance of nonlinearity for memory under nonequilibrium settings.
The impact of active noise on memory, as described by $H_{\text{eff}}$
of Eq. (\ref{eq:heff}), are summarized in Fig. \ref{fig:h-eff}.
The limit of passive noise is recovered by setting $\tau=0$, which
yields the original Hamiltonian {[}Eq. (\ref{eq:h}){]}. 

Note that the ``active stabilization'' described by $H_{\text{eff}}$
of Eq. (\ref{eq:heff}) arises from the $-\tau T\nabla_{\bm{\theta}}^{2}H$
term of the general form of $H_{\text{eff}}$ {[}Eq. (\ref{eq:general-heff}){]}
rather than the $\frac{\tau}{2}\left|\nabla_{\bm{\theta}}H\right|^{2}$
term. The former energy correction, which depends on the curvature
of the potential energy surface, has been shown to stabilize certain
particle configurations in a related system, whose couplings go as
the cosine of relative orientations \citep{maitra_enhanced_2020}.
The result here stands in contrast to our previous work \citep{behera_enhancing_2022}
on the continuous-spin Hopfield network of \citep{bolle_spherical_2003}
under active noise: there, the interactions scale with the number
$N\gg1$ of spins in such a way that the the $-\tau T\nabla_{\bm{\theta}}^{2}H$
term is negligible compared to the $\frac{\tau}{2}\left|\nabla_{\bm{\theta}}H\right|^{2}$
term. 

\subsection{Replica calculation\label{subsec:Replica-theory}}

Next, we apply replica theory \citep{mezard_spin_1987} to the effective
Hamiltonian $H_{\text{eff}}$ of Eq. (\ref{eq:heff}). Details of
the calculation are presented in Appendix \ref{sec:appendix-replica}.
Replica theory allows one to calculate order parameters, such as the
overlap with a pattern, of a system in the thermodynamic limit ($N\rightarrow\infty$)
and averaged over randomly generated parameters of the Hamiltonian
(known as ``quenched disorder'' in the literature \citep{amit_storing_1985}),
such as the learned patterns. From the order parameters, one can determine
what phase the system is in, e.g., ferromagnetic (retrieval) and spin
glass (no retrieval). Following the convention of \citep{amit_statistical_1987,amit_storing_1985,coolen_bookchapter1_2001},
a retrieval phase is defined by nonzero overlap with the target pattern
(Section \ref{subsec:phases}). In contrast, our simulated results
above (Figs. \ref{fig:passive}-\ref{fig:active}) use the criterion
that the overlap exceeds 0.8 (caption of Fig. \ref{fig:passive}).
Thus, our analysis below is limited to qualitative comparisons between
the phase diagrams obtained by replica theory and simulation.
Furthermore, our replica calculation is limited to $T > 0.1$;
for $T \rightarrow 0$, we anticipate that the saddle-point equations [Eqs. (\ref{eq:m})-(\ref{eq:r})] 
which determine the order parameters can yield unstable solutions (Sec. \ref{subsec:replica-symmetric}). 

\begin{figure}
\centering\includegraphics[width=3.3in]{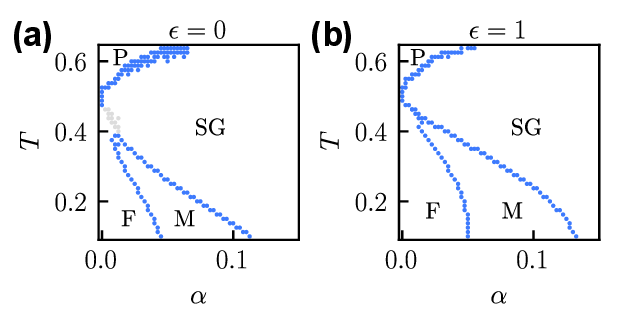}

\caption{Retrieval phase diagram under passive noise, as calculated by theory,
for stability-enhancing coupling (a) $\epsilon=0$ and (b) $\epsilon=1$.
In the ferromagnetic (F) retrieval phase, the global minimum of free
energy function $f$ {[}Eq. (\ref{eq:fred}){]} has nonzero overlap
with the target pattern. In the mixed (M) retrieval phase, a local
(but not global) minimum of $f$ has nonzero overlap with the target
pattern. No retrieval occurs in the spin glass (SG) and paramagnetic
(P) phases, in which all minima of $f$ have zero overlap with the
target pattern. The gray points indicate where no minima of $f$ could
be found numerically (see Appendix \ref{subsec:replica-numerics}).
Note that the plots start at $T = 0.1$, 
in contrast to the corresponding results from simulations 
(Fig. \ref{fig:passive}).
\label{fig:replica-passive}}
\end{figure}

We first consider the system under passive noise, for which $H_{\text{eff}}$
reduces to $H$ {[}Eq. (\ref{eq:h}){]}, the original Hamiltonian.
To our knowledge, $H$ has not been previously studied using replica
theory. Figs. \ref{fig:replica-passive}a and \ref{fig:replica-passive}b
show the phase diagram calculated by this method for $\epsilon=0$
and $\epsilon=1$, respectively. Both plots qualitatively resemble
the corresponding phase diagram for the Hopfield model \citep{amit_statistical_1987,amit_storing_1985,coolen_bookchapter1_2001}.
At $T\lesssim0.5$, increasing pattern loading induces transitions
from the ferromagnetic retrieval phase (i.e., a global free energy
minimum has significant overlap with the target pattern) to the mixed
retrieval phase (i.e., the global minimum becomes a local minimum)
to the spin glass phase, which does not support retrieval. The same
phase transitions can be triggered by increasing $T$. At $T\gtrsim0.5$,
the target pattern cannot be retrieved at all; the system is in the
paramagnetic phase at low pattern loading and the spin glass phase
at high pattern loading. 

Consistent with simulations (Fig. \ref{fig:passive}), replica theory
(Fig. \ref{fig:replica-passive}) reveals that the stability-enhancing
term ($\epsilon>0$) raises the critical temperature and pattern loading
at which the system transitions from a retrieval (mixed) to a no-retrieval
(spin glass) phase. Also increased are the critical values separating
the two retrieval phases.

\begin{figure}
\centering\includegraphics[width=3.3in]{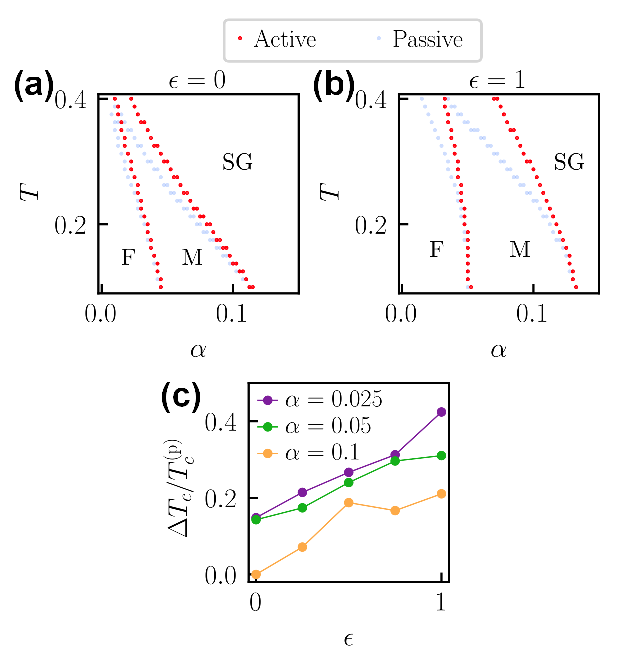}

\caption{(a, b) Retrieval phase diagram under active noise ($\tau=0.2$, red),
as calculated by theory, for stability-enhancing coupling (a) $\epsilon=0$
and (b) $\epsilon=1$. The dotted lines indicate where phase transitions
occur under passive noise (compare to corresponding panels of Fig.
\ref{fig:replica-passive}). See the caption of Fig. \ref{fig:replica-passive}
for additional details. (c) Relative difference ($\Delta T_{c}/T_{c}^{(\text{p})}$,
where $\Delta T_{c}=T_{c}^{(\text{a})}-T_{c}^{(\text{p})}$) in the
critical noise strength (for the $\text{M}\rightarrow\text{SG}$ transition)
between active ($T_{c}^{(\text{a})}$) and passive ($T_{c}^{(\text{p})}$)
noise, as calculated by theory and as a function of $\epsilon$. Results are shown for various
pattern loadings ($\alpha$). \label{fig:replica-active}}
\end{figure}

We proceed to the corresponding phase diagrams for the system under
active noise. Since effective equilibrium theory is correct to $O(\tau)$
(Section \ref{subsec:effective-equilibrium}), our application of
replica theory to $H_{\text{eff}}$ of Eq. (\ref{eq:heff}) assumes
$\tau=0.2$: this value is small enough for the theory to be a reasonable
approximation but large enough to observe appreciable differences
between active and passive noise. Fig. \ref{fig:replica-active}a-b
shows phase diagrams obtained by the theory. Replacing passive with
active noise enables the target pattern to be retrieved at higher
noise strengths (for a given pattern loading) and pattern loadings
(for a given noise strength). In addition, the relative increase in
the critical noise strengths scales with $\epsilon$ (Fig. \ref{fig:replica-active}c).
Remarkably, these features also appear in the simulated phase diagrams
(Fig. \ref{fig:active}, a-b, d) for $\tau=1$, which falls outside
the range of persistence times ($\tau\ll1)$ for which effective equilibrium
theory is expected to be valid. This agreement suggests that $H_{\text{eff}}$
of Eq. (\ref{eq:heff}) can qualitatively describe how active noise
improves memory, even for $\tau\geq1$. As discussed in Section \ref{subsec:effective-equilibrium},
the impact of active noise in this effective picture is to strengthen
the interactions that stabilize the patterns, where the interactions
with higher nonlinearity are enhanced over those with less. The replica
calculation further reveals that active noise delays the transition
(with respect to increasing noise strength and pattern loading) from
the ferromagnetic retrieval phase to the mixed retrieval phase. 

Before concluding, let us briefly discuss the quantitative aspects
of the theory. In Appendix \ref{subsec:overlap}, we evaluate how
well the theory can reproduce the final overlap computed by simulations.
For the case of passive noise, the agreement is semi-quantitative.
For active noise, we observe similar agreement between the theory
and simulations with the same persistence time ($\tau=0.2$), though
only at relatively low $T$. More details can be found in Appendix
\ref{subsec:overlap}.

\section{Conclusion \label{sec:conclusion}}

We studied how active noise impacts periodic oscillators whose couplings
endow the system with associative memory. One type of interaction
encodes patterns, while the second type, a more nonlinear interaction,
enhances the stability of the patterns. Through simulations, we showed
that pattern retrieval under active noise is more robust to noise
strength and number of learned patterns than under passive noise.
This improvement in memory scales with the strength of the more nonlinear
coupling relative to the less nonlinear coupling. 

Using effective equilibrium theory, which is perturbative with respect
to the persistence time of active noise, we analytically derived an
effective Hamiltonian {[}Eq. (\ref{eq:heff}){]} that captures how
the temporal correlations of active noise resculpt the potential energy
landscape and thereby improve memory. Specifically, active noise leaves
the oscillator interactions unchanged in form but increased in strength,
with the more nonlinear interaction being preferentially enhanced.
As a result of the stronger couplings, the energy basins at the pattern
configurations become deeper. Unlike the Hopfield model studied in
\citep{bolle_spherical_2003,behera_enhancing_2022}, the effect of
active noise here arises mainly from the curvature of the original
energy surface rather than the slope.

We then demonstrated qualitative agreement between this effective
picture and the simulated results by applying replica theory to the
effective Hamiltonian. This agreement is remarkable because effective
equilibrium theory strictly holds for persistence time $\tau\ll1$
while the simulations were run with $\tau=1$. In addition, the replica
analysis reveals that replacing passive with active noise can increase
the critical noise strength at which a target pattern is converted
from a global minimum of the (effective) free energy to a local minimum. 

Overall, our findings highlight that having nonlinear couplings between
neurons can be essential for associative memory under nonequilibrium
conditions and thus biologically relevant contexts. For the Hopfield
model in the absence of noise, highly nonlinear interactions have
been shown to drastically increase the critical number of learned
patterns beyond which the patterns can no longer be retrieved \citep{krotov_dense_2016,ramsauer_hopfield_2021}.
Our study suggests that such nonlinearities may confer additional
benefits to memory in noisy and nonequilibrium environments. 

\begin{acknowledgments}
M. D. and S. V. are supported by DOE BES Grant No. DE-SC0019765.
A. K. B. is supported by a fellowship from 
the Department of Chemistry at the University of Chicago. 
We thank Carlos Floyd, David Martin, and Aditya Nandy 
for their feedback on earlier drafts of the manuscript.

\end{acknowledgments}

\appendix

\section{Details of replica calculation\label{sec:appendix-replica}}

In this section, we apply replica theory \citep{mezard_spin_1987}
to effective Hamiltonian $H_{\text{eff}}$ of Eq. (\ref{eq:heff}).
Our replica calculation largely follows those \citep{amit_statistical_1987,coolen_bookchapter1_2001}
for the Hopfield Hamiltonian. Note that the first term of $H_{\text{eff}}$
{[}Eq. (\ref{eq:heff}){]} has the same form as the Hamiltonian of
the Hopfield model. 

\subsection{Gauge fixing}

To simplify the calculation, we first choose a gauge for the patterns.
Without loss of generality, we arbitrarily designate pattern 1 as
the target pattern, $\bm{\xi}\equiv\bm{\xi}^{1}$, as we have done
in our numerical simulations (Sec. \ref{sec:simulations}). We also
apply an appropriate rotation to all memories such that $\bm{\xi}=(1,\dots,1)$.
This transformation, which is orthogonal, does not change the final
result of the calculation. Since we have fixed the target pattern,
we only need to average over the other $p-1$ patterns. 

\subsection{Replica theory}

In replica theory \citep{mezard_spin_1987}, the free energy per oscillator---at
(effective) equilibrium, averaged over the non-target patterns, and
in the thermodynamic limit ($N\rightarrow\infty$)---is expressed
in terms of a ``replicated'' partition function: 
\begin{equation}
\lim_{N\rightarrow\infty}\frac{\overline{F}}{N}=-\lim_{N\rightarrow\infty}\frac{1}{\beta N}\overline{\ln Z}=-\lim_{N\rightarrow\infty}\lim_{n\rightarrow0}\frac{1}{\beta Nn}\ln\overline{Z^{n}},\label{eq:fbn}
\end{equation}
where $\overline{\cdots}$ denotes the pattern average, $\beta=1/T$
($k_{B}=1$) is the inverse temperature, $Z=\int d\bm{\theta}\exp\left[-\beta H_{\text{eff}}(\bm{\theta})\right]$
is the partition function, and $n$ is the number of replicas of the
system. Due to the periodicity of $H_{\text{eff}}$, the integral
over each $\theta_{i}$ ($i=1,\dots,N$) can be chosen to run over
any (contiguous) interval of length $2\pi$ (e.g., $[0,2\pi]$). Plugging
in $H_{\text{eff}}$ {[}Eq. (\ref{eq:heff}){]} of the chosen gauge,
the replicated partition function can be written as 
\begin{align}
\overline{Z^{n}} & =\int d\overleftrightarrow{\bm{\theta}}\underbrace{\exp\left\{ \frac{\beta\zeta_{1}}{2N}\sum_{\gamma}\left[\bm{\xi}\cdot\mathbf{c}(\bm{\theta}^{\gamma})\right]^{2}\right\} }_{\equiv E_{1}}\nonumber \\
 & \quad\times\underbrace{\overline{\exp\left\{ \frac{\beta\zeta_{1}}{2N}\sum_{\gamma}\sum_{\mu>1}\left[\bm{\xi}^{\mu}\cdot\mathbf{c}(\bm{\theta}^{\gamma})\right]^{2}\right\} }}_{\equiv E_{2}}\nonumber \\
 & \quad\times\underbrace{\exp\left\{ \frac{\beta\zeta_{2}\epsilon}{4N}\sum_{\gamma}\left[\mathbf{1}\cdot\mathbf{c}(2\bm{\theta}^{\gamma})\right]^{2}\right\} }_{\equiv E_{3}},\label{eq:zn}
\end{align}
where $\overleftrightarrow{\bm{\theta}}=(\bm{\theta}^{1},\dots,\bm{\theta}^{n})$
contains the configurations $\bm{\theta}^{\gamma}=(\theta_{1}^{\gamma},\dots,\theta_{N}^{\gamma})$
of replica $\gamma=1,\dots,n$. Note that $E_{2}$ includes the pattern
average ($\overline{\cdots}$). We next write each $E_{i}$ as an
integral, whose integration variables will become order parameters.
The terms $E_{1}$ and $E_{2}$ are the contributions from the target
pattern and non-target patterns, respectively. In analogy to the replica
calculation for the Hopfield model \citep{amit_statistical_1987,coolen_bookchapter1_2001},
we have
\begin{align}
E_{1} & =\left(\frac{N\beta\zeta_{1}}{2\pi}\right)^{n/2}\nonumber \\
 & \quad\times\int d\mathbf{m}\exp\beta\zeta_{1}\left\{ \sum_{\gamma}m^{\gamma}\left[\bm{\xi}\cdot\mathbf{c}(\bm{\theta}^{\gamma})\right]-\frac{N|\mathbf{m}|^{2}}{2}\right\} ,\label{eq:e1}\\
E_{2} & =\left(\frac{N}{2\pi}\right)^{n^{2}}\nonumber \\
 & \quad\times\int d\mathbf{q}d\hat{\mathbf{q}}\exp\left\{ i\sum_{\gamma\kappa}\hat{q}^{\gamma\kappa}\left[Nq^{\gamma\kappa}-\mathbf{c}(\bm{\theta}^{\gamma})\cdot\mathbf{c}(\bm{\theta}^{\kappa})\right]\right\} \nonumber \\
 & \quad\times\left[\det\left(\mathbf{I}-\beta\zeta_{1}\mathbf{q}\right)\right]^{-p/2},\label{eq:e2}
\end{align}
where $\mathbf{I}$ is the $n\times n$ identity matrix. In tensor
notation, the integration variables are $m^{\gamma}$, $q^{\gamma\kappa}$,
and $\hat{q}^{\gamma\kappa}$ for $\gamma,\kappa=1,\dots,n$. Note
that Eq. (\ref{eq:e2}) has been written assuming $N\rightarrow\infty$
\citep{amit_statistical_1987,coolen_bookchapter1_2001}. The term
$E_{3}$ is the contribution from the stability-enhancing term {[}second
term of Eq. (\ref{eq:h}){]}. Using a standard Gaussian integral identity,
it is straightforward to show that
\begin{align}
E_{3} & =\left(\frac{N\beta\zeta_{2}\epsilon}{4\pi}\right)^{n/2}\nonumber \\
 & \times\int d\mathbf{w}\exp\frac{\beta\zeta_{2}\epsilon}{2}\left\{ \sum_{\gamma}w^{\gamma}\left[\mathbf{1}\cdot\mathbf{c}(2\bm{\theta}^{\gamma})\right]-\frac{N|\mathbf{w}|^{2}}{2}\right\} ,\label{eq:e3}
\end{align}
where the integration variables are (in tensor notation) $w^{\gamma}$
for $\gamma=1,\dots,n$. We can then substitute Eqs. (\ref{eq:e1})-(\ref{eq:e3})
into Eq. (\ref{eq:zn}) and write the replicated partition function
in the form $\overline{Z^{n}}=\int d\mathbf{m}d\mathbf{w}d\mathbf{q}d\hat{\mathbf{q}}\exp\left(\cdots\right)$.
Finally, we evaluate the free energy per oscillator {[}Eq. (\ref{eq:fbn}){]}
using the saddle-point method,
\begin{equation}
\lim_{N\rightarrow\infty}\frac{\overline{F}}{N}=\min_{\left(\mathbf{m},\mathbf{w},\mathbf{q},\hat{\mathbf{q}}\right)}f\left(\mathbf{m},\mathbf{w},\mathbf{q},\hat{\mathbf{q}}\right),\label{eq:fbn-min}
\end{equation}
where
\begin{align}
f\left(\mathbf{m},\mathbf{w},\mathbf{q},\hat{\mathbf{q}}\right) & =\lim_{n\rightarrow0}\frac{1}{\beta n}\left[\frac{\beta\zeta_{1}|\mathbf{m}|^{2}}{2}+\frac{\beta\zeta_{2}\epsilon|\mathbf{w}|^{2}}{4}\right.\nonumber \\
 & \quad-i\sum_{\gamma\kappa}\hat{q}^{\gamma\kappa}q^{\gamma\kappa}+\frac{\alpha}{2}\ln\det\left(\mathbf{I}-\beta\zeta_{1}\mathbf{q}\right)\nonumber \\
 & \quad\left.-\frac{1}{N}\ln\int d\overleftrightarrow{\bm{\theta}}G\left(\mathbf{m},\mathbf{w},\mathbf{q},\hat{\mathbf{q}},\overleftrightarrow{\bm{\theta}}\right)\right]\label{eq:f}
\end{align}
is the function to be minimized and
\begin{align}
G\left(\mathbf{m},\mathbf{w},\mathbf{q},\hat{\mathbf{q}},\overleftrightarrow{\bm{\theta}}\right) & =\exp\left\{ \beta\zeta_{1}\sum_{\gamma}m^{\gamma}\left[\bm{\xi}\cdot\mathbf{c}(\bm{\theta}^{\gamma})\right]\right.\nonumber \\
 & \quad+\frac{\beta\zeta_{2}\epsilon}{2}\sum_{\gamma}w^{\gamma}\left[\mathbf{1}\cdot\mathbf{c}(2\bm{\theta}^{\gamma})\right]\nonumber \\
 & \quad\left.-i\sum_{\gamma\kappa}\hat{q}^{\gamma\kappa}\left[\mathbf{c}(\bm{\theta}^{\gamma})\cdot\mathbf{c}(\bm{\theta}^{\kappa})\right]\right\} .\label{eq:G}
\end{align}

As mentioned above, the independent variables of $f$ correspond to
order parameters. The parameter values that minimize $f$ characterize
the minima of the free energy landscape. In particular, the values
that globally minimize $f$ correspond to the (effective) equilibrium
values and yield the (effective) equilibrium free energy through Eq.
(\ref{eq:fbn-min}). 

The physical meaning of the order parameters can be deduced from the
saddle-point equations. For example, the equations obtained by extremizing
$f$ {[}Eq. (\ref{eq:f}){]} with respect to $m^{\gamma}$ and $\hat{q}^{\gamma\kappa}$
are
\begin{align}
m^{\gamma} & =\frac{1}{N}\left\langle \bm{\xi}\cdot\mathbf{c}(\bm{\theta}^{\gamma})\right\rangle ,\\
q^{\gamma\kappa} & =\begin{cases}
\frac{1}{N}\sum_{i=1}^{N}\left\langle \cos^{2}\theta_{i}^{\gamma}\right\rangle , & \gamma=\kappa,\\
\frac{1}{N}\sum_{i=1}^{N}\left\langle \cos\theta_{i}^{\gamma}\cos\theta_{i}^{\kappa}\right\rangle , & \gamma\neq\kappa,
\end{cases}
\end{align}
where we have defined the average $\left\langle h\left(\overleftrightarrow{\bm{\theta}}\right)\right\rangle =\frac{\int d\overleftrightarrow{\bm{\theta}}G\left(\overleftrightarrow{\bm{\theta}}\right)h\left(\overleftrightarrow{\bm{\theta}}\right)}{\int d\overleftrightarrow{\bm{\theta}}G\left(\overleftrightarrow{\bm{\theta}}\right)}$
for any function $h\left(\overleftrightarrow{\bm{\theta}}\right)$.
As in the Hopfield model \citep{amit_statistical_1987,coolen_bookchapter1_2001},
$m^{\gamma}$ represents the overlap with the target memory (see Sec.
\ref{sec:Model} for definition of overlap), and $q^{\gamma\kappa}$
for $\gamma\neq\kappa$ is the Edwards-Anderson parameter \citep{edwards_theory_1975}
except with the variable of spin $i$ replaced by $\cos\theta_{i}$.
Thus, minimization of $f$ will provide information about the overlap
at free energy minima.

\subsection{Replica-symmetric solution \label{subsec:replica-symmetric}}

To minimize $f$ explicitly, we must impose some constraints on the
values of the order parameters. Let us focus on the replica-symmetric
ansatz 
\begin{align}
m^{\gamma} & =m,\\
w^{\gamma} & =w,\\
q^{\gamma\kappa} & =Q\delta_{\gamma\kappa}+q(1-\delta_{\gamma\kappa}),\\
\hat{q}^{\gamma\kappa} & =\frac{i\alpha(\beta\zeta_{1})^{2}}{2}\left[R\delta_{\gamma\kappa}+r(1-\delta_{\gamma\kappa})\right].
\end{align}
Then $f$ {[}Eq. (\ref{eq:f}){]} reduces to
\begin{align}
f(m,w,Q,q,R,r) & =\frac{\zeta_{1}m^{2}}{2}+\frac{\zeta_{2}\epsilon}{2}\left(\frac{w^{2}}{2}+w\right)\nonumber \\
 & \quad+\frac{\alpha}{2}\Biggl\{\beta\zeta_{1}^{2}\left(RQ-rq\right)\nonumber \\
 & \quad+\frac{1}{\beta}\ln\left[1-\beta\zeta_{1}(Q-q)\right]\nonumber \\
 & \quad-\frac{\zeta_{1}q}{1-\beta\zeta_{1}(Q-q)}\Biggr\}\nonumber \\
 & \quad-\frac{1}{\beta}\int Dz\ln\mathcal{I}_{0}(m,w,R,r,z),\label{eq:fred}
\end{align}
where
\begin{align}
\mathcal{I}_{v}(m,w,R,r,z) & =\int d\theta g(m,w,R,r,\theta,z)\cos^{v}\theta,\\
g(m,w,R,r,\theta,z) & =\exp\beta\Biggl\{\zeta_{1}\left(m+z\sqrt{\alpha r}\right)\cos\theta\nonumber \\
 & \quad+\left[\zeta_{2}\epsilon w+\frac{\alpha\beta\zeta_{1}^{2}}{2}(R-r)\right]\cos^{2}\theta\Biggr\}.
\end{align}
In evaluating the determinant of Eq. (\ref{eq:f}), we used the fact
that $\mathbf{q}$ is a $n\times n$ matrix of the form $M_{ij}=a\delta_{ij}+b(1-\delta_{ij})$
and thus has eigenvalues $Q-q$ and $Q+(n-1)q$ with multiplicities
$n-1$ and 1, respectively. In addition, the integral in Eq. (\ref{eq:f})
was simplified to that in Eq. (\ref{eq:fred}) using the following
facts: $\bm{\xi}=(1,\dots,1)$, all oscillators are identical, all
replicas are identical, and
\begin{equation}
\lim_{n\rightarrow0}\frac{\ln\int Dz\,h^{n}(z)}{n}=\int Dz\,\ln h(z)
\end{equation}
for any function $h$ such that the limit can be evaluated using L'H\^opital's
rule. The extrema of $f$ are solutions of the saddle-point equations
\begin{align}
m & =\int Dz\,u_{1}(m,w,R,r,z),\label{eq:m}\\
w & =-1+2Q,\label{eq:w}\\
Q & =\int Dz\,u_{2}(m,w,R,r,z),\\
q & =\int Dz\,u_{1}^{2}(m,w,R,r,z),\\
R & =\frac{1}{\beta\zeta_{1}}\frac{1-\beta\zeta_{1}(Q-2q)}{[1-\beta\zeta_{1}(Q-q)]^{2}},\label{eq:R}\\
r & =\frac{q}{[1-\beta\zeta_{1}(Q-q)]^{2}},\label{eq:r}
\end{align}
where
\begin{align}
u_{v}(m,w,R,r,z) & =\frac{\mathcal{I}_{v}(m,w,R,r,z)}{\mathcal{I}_{0}(m,w,R,r,z)}.
\end{align}

Given that $H_\text{eff}$ [Eq. (\ref{eq:heff})] is similar in form to the Hamiltonian of the Hopfield model \citep{hopfield_neural_1982}, 
we expect the solution(s) of Eqs. (\ref{eq:m})-(\ref{eq:r}) to eventually become unstable as $T \rightarrow 0$
for some values of $\alpha$.
Indeed, such instability in the replica-symmetric solution 
was found for the Hopfield model
\citep{amit_statistical_1987,amit_storing_1985,naef_reetrant_1992,coolen_bookchapter1_2001}.
This issue can be addressed by using an ansatz 
that breaks replica symmetry \citep{mezard_spin_1987}, 
which has been done for the Hopfield model 
\citep{crisanti_saturation_1986,steffan_replica_1994}.

\subsection{Numerical details\label{subsec:replica-numerics}}

To numerically solve the saddle-point equations (\ref{eq:m})-(\ref{eq:r}),
we optimize with respect to $m$, $Q$, and $q$ only, while $w$,
$Q$, and $q$ are constrained by Eqs. (\ref{eq:w}), (\ref{eq:R}),
and (\ref{eq:r}), respectively. The computation is done in Python
using the \texttt{optimize.root} function (all keyword parameters
take their default value, including \texttt{method='hybr'}) of the
SciPy package. We repeat the calculation for initial conditions $(m,Q,q)=$
$(0.9,0.9,0.9)$, $(0,0.99,0.9)$, $(0,0.9,0.9)$, $(0,0.8,0.6)$,
$(0,0.6,0)$, $(0,0.4,0.1)$, $(0,0.49999999,0)$. We accept solutions
$(m,Q,q)$ for which the maximum relative error, with respect to the
nonzero elements, is less than $10^{-6}$. For each order parameter
$x=m,Q,q$, the relative error is calculated as $(x_{\text{solution}}-x_{\text{actual}})/x_{\text{solution}}$,
where $x_{\text{solution}}$ is the value in the solution and $x_{\text{actual}}$
is the value computed from the right-hand side of the saddle-point
equation for $x$. We note that, for some choices of model parameters
($\alpha$, $\epsilon$, $T$, $\tau$), no solution could be found
using the attempted initial conditions. 

\subsection{Determination of phases\label{subsec:phases}}

For a given choice of model parameters ($\alpha$, $\epsilon$, $T$,
$\tau$), we assign a phase according to the sets of order parameters
that solve saddle-point equations (\ref{eq:m})-(\ref{eq:r}) and
thus extremize $f$ {[}Eq. (\ref{eq:fred}){]}. Specifically, the
phase is determined by $(m,q)$ of the global minimum of $f$ and
$m$ of the local minima, as done in replica calculations of the Hopfield
model \citep{amit_statistical_1987,amit_storing_1985,coolen_bookchapter1_2001}
and reviewed in Table \ref{tab:phases}. The existence of a local
minimum with $m\neq0$ implies that the target pattern can be retrieved,
while the absence thereof implies that retrieval is not possible.

\begin{table*}
\centering%
\begin{tabular}{>{\centering}p{1.2in}>{\centering}p{1.2in}>{\centering}p{1.6in}>{\centering}p{1.6in}}
\hline 
Phase & Retrieval 

or no retrieval

of target pattern? & $(m,q)$ 

of the global minimum 

of $f$ & $m$ 

of the local minima (if any) 

of $f$\tabularnewline
\hline 
ferromagnetic (F)  & retrieval & $m\neq0$, $q\neq0$ & $m=$ any value\tabularnewline
mixed (M) & retrieval & $m=0$, $q\neq0$ & $m\neq0$ for $\geq1$ minimum\tabularnewline
spin glass (SG) & no retrieval & $m=0$, $q\neq0$ & $m=0$ for all minima\tabularnewline
paramagnetic (P) & no retrieval & $m=0$, $q=0$  & $m=0$ for all minima\tabularnewline
\hline 
\end{tabular}

\caption{Phases of Hopfield-like models as described within replica theory.\label{tab:phases}}

\end{table*}

\subsection{Calculation of overlap: theory versus simulation\label{subsec:overlap}}

\begin{figure*}
\centering\includegraphics[width=6.6in]{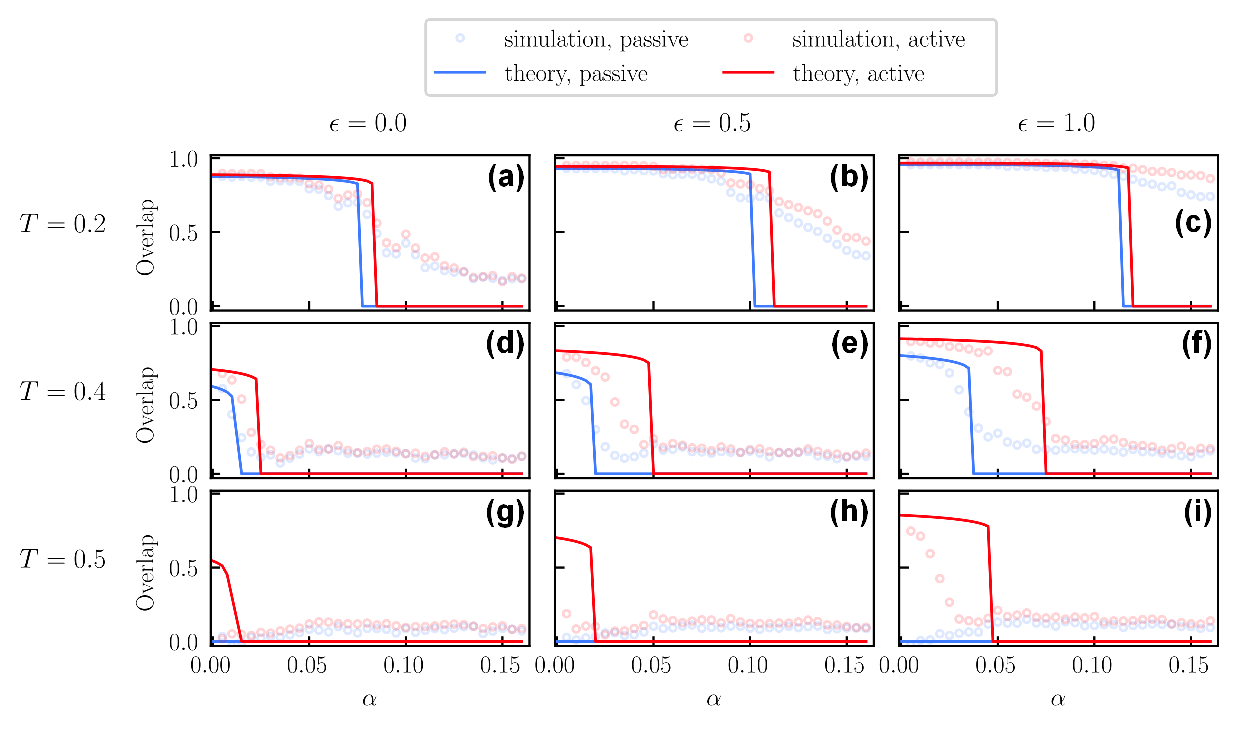}

\caption{Overlap with target pattern versus pattern loading $\alpha$, as calculated
by simulation (circles) and theory (lines). Results are shown for
passive (purple) and active (blue, $\tau=0.2$) noise; noise strength
(a-c) $T=0.2$, (d-f) $T=0.4$, and (g-i) $T=0.5$; and stability-enhancing
coupling (a, d, g) $\epsilon=0$, (b, e, h) $\epsilon=0.5$, and (c,
f, i) $\epsilon=1$.\label{fig:simulation-vs-theory}}
\end{figure*}

The highest overlap, with respect to all minima of $f$, is expected
to correspond to the final overlap in a simulation of pattern retrieval.
Fig. \ref{fig:simulation-vs-theory} compares the highest overlap
calculated in our theory (lines) with the final overlap obtained in
our simulations (circles). 

We first focus on the system under passive noise (Fig. \ref{fig:simulation-vs-theory},
blue). Quantitative agreement is obtained at low pattern loading,
while the theoretical overlap reasonably approximates the decay of
the simulated overlap as the pattern loading increases. Note, however,
that the theory overestimates the overlap during the decay and incorrectly
predicts that the overlap approaches zero. We attribute these discrepancies
to the fact that the simulations are done for a finite number $N$
of oscillators while the theory assumes $N\rightarrow\infty$. 

Having validated the theory for the case of passive noise, we now
turn to the system under active noise with $\tau=0.2$ (Fig. \ref{fig:simulation-vs-theory},
red). Like the case of passive noise, we see good agreement between
simulation and theory at relatively low pattern loading and noise
strength. In contrast, at sufficiently high noise strength, the theory
predicts significant overlap under active noise while the simulations
show little overlap (Figs. \ref{fig:simulation-vs-theory}g-\ref{fig:simulation-vs-theory}h).
We believe this discrepancy is related to the following: the effective
Hamiltonian {[}Eq. (\ref{eq:heff}){]} used in the replica calculation
does not include the $\frac{\tau}{2}\left|\nabla_{\bm{\theta}}H\right|^{2}$
term of the full effective Hamiltonian {[}Eq. (\ref{eq:general-heff}){]}
derived by effective equilibrium theory (Section \ref{subsec:effective-equilibrium}).
At higher noise strengths, the system is kicked away from the patterns,
and so $-\nabla_{\bm{\theta}}H$, the force pushing the system towards
the patterns, becomes significant.

\bibliographystyle{apsrev4-2}
%

\end{document}